# Aqueous humor dynamics in human eyes: a lattice Boltzmann study

**Zhangrong Qin[1], Lingjuan Meng[1], Fan Yang[2], Chaoying Zhang[1,]\* & Binghai Wen[1,\*]**

[1] Guangxi Key Lab of Multi-source Information Mining & Security, Guangxi Normal University, Guilin 541004, China

[2] Ophthalmology Department, Nanxishan Hospital of Guangxi Zhuang Autonomous Region, Guilin 541002, China

**\* Correspondence:** Email: Zhangcy@gxnu.edu.cn (C. Zhang); Oceanwen@gxnu.edu.cn (B. Wen).

**Abstract:** This paper presents a lattice Boltzmann model to simulate the aqueous humor (AH) dynamics in the human eyes by involving incompressible Navier-Stokes flow, heat convection and diffusion, and Darcy seepage flow. Verifying simulations indicate that the model is stable, convergent and robust. Further investigations were carried out, including the effects of heat convection and buoyancy, AH production rate, permeability of trabecular meshwork, viscosity of AH and anterior chamber angle on intraocular pressure (IOP). The heat convection and diffusion can significantly affect the flow patterns in the healthy eye, and the IOP can be controlled by increasing the anterior chamber angle or decreasing the secretion rate, the drainage resistance and viscosity of AH. However, the IOP is insensitive to the viscosity of the AH, which may be one of the causes that the viscosity would not have been considered as a factor for controlling the IOP. It's interesting that all these factors have more significant influences on the IOP in pathologic eyes than healthy ones. The temperature difference and the eye-orientation have obvious influence on the cornea and iris wall shear stresses. The present model and simulation results are expected to provide an alternative tool and theoretical reference for the study of AH dynamics.

## 1. Introduction

Glaucoma is a severe progressive ocular disease, which can result in vision loss and is the 2rd most common cause of blindness worldwide [1]. As elderly population grows, more and more people are expected to be affected by this disease. For a long time, prophylaxis and treatment of glaucoma have attracted substantial attentions of the researchers all over the world. Lots of researches have shown that glaucoma is closely related with the dynamics of aqueous humor (AH) [2]. The anterior segment of eye is filled with AH, a clear, colorless, water-like fluid that is continuously secreted by the ciliary body (CB), located just posterior to the iris. It passes through the gap between the iris and the lens into the anterior chamber (AC), and then circulates there due to natural convection. Eventually, the bulk of AH drains from the eye via specialized tissues consisted of the trabecular meshwork (TM), the juxtacanalicular connective tissue (JCT), the endothelial lining of Schlemm's canal (SC), the collecting channels and the aqueous veins [3]. As these specialized tissues can cause a significant hydrodynamic flow resistance, a positive pressure in the eye is required to drain AH out of the eye, and the positive pressure is called intraocular pressure (IOP). IOP takes a central role in the AH dynamics, which stabilizes the otherwise flaccid eye and ensures its normal physiological functions. When an obstruction of TM or SC occurs, an elevation in IOP is then observed, and this direct mechanical effect is the most common risk factor for glaucoma progression. High-level of IOP, sustained for a long time, can damage the optic nerve in the eye and eventually result in vision loss, which is the primary cause of blindness in glaucoma. Fortunately, several studies have shown evidence that IOP can be controlled by decreasing the production of AH and/or decreasing the hydrodynamic resistance of the trabecular outflow pathways for the drainage of AH [4-6]. Therefore revealing the pathogenesis and development mechanisms of glaucoma, in terms of the dynamics of AH, is crucial to its prophylaxis and therapy. Over the years, substantial efforts have been devoted to better understanding the dynamics of AH, both analytically and numerically [2,7].

Glaucoma is often, although not always, accompanied by an elevation in IOP, and this elevation is usually known to be resulted from an increase in the hydrodynamic resistance to AH drainage. Originally, lots of work focused on the analysis of TM as a porous material filled with a gel-like biopolymer [8], and the predictions on the overall permeability of TM by this method is quite close to the experimental data. However, the subsequent treatments of TM, according to the predictions of this model, showed that the effect of reducing hydraulic resistance is much smaller than would be expected [9]. In another hand, it is also important for the mechanism of AH, which drains from the eye via the small pores in the cellular lining of Schlemm's canal [10]. Assumed flow through a single pore and modeled the whole cellular lining of SC as a parallel network consisting of hydrodynamically non-interacting pores, the resistance of a single pore to AH outflow can be estimated by Sampson's theory in low-Reynolds number hydrodynamics. However, the results are much lower than the observed total resistance of aqueous outflow [2]. Johnson et al. [11] noted a fact that there is a hydrodynamic interaction between the upstream porous material of TM and the endothelial lining of SC. Based on this fact, a simplified unit-cell model, composing of a single pore and the upstream region of the porous medium drained by that pore, was proposed, with which the calculated hydraulic resistance of the ensemble structure is in agreement with experimental measurements [12]. By using the finite element method, Bradley & Heys [13] researched the effects of variable permeability on AH outflow, and found that the overall resistance of JCT would be increased when the tissue is assumed to be heterogeneous (instead of homogenous).

In addition to the drainage of AH from TM, the overall flow process of AH in human eye, involving AH generation, flow, circulation, and finally drainage, attracts researchers' attentions all the way. By using analytic method, Avtar & Srivastava [14] constructed a simple mathematical model by a series of approximations, and discussed the natural convectional flow of AH in AC of a human eye. Considering the coupling of Navier-Stokes and Darcy flows, Crowder & Ervin [15] proposed a numerical method for investigating the fluid pressure in human eye without the temperature difference across AC. In order to investigate the drug delivery through the cornea, from a therapeutic lens to AC of eye, Ferreira et.al [6] established a mathematic model for the AH flow in both healthy and pathologic eyes, and numerically studied the interaction of drug flow with the dynamics of AH. In consideration of the existence of segmental outflow, Loke et.al [16] developed a numerical model for investigating the influences of the segmental AH outflow and the eye orientation on ocular drug delivery, and they suggested the design of ophthalmic drug delivery is needed to re-evaluate according to these effects. Using the method of theoretical analysis combined with numerical calculation, Fitt & Gonzalez studied in detail the flows of AH in AC driven by various physical mechanisms [17]. Zhang et.al [18] proposed a 3D computational model for drug delivery in anterior eye after subconjunctival and episcleral implantation, and found that subconjunctival implantation is more effective than episcleral implantation. In addition, Canning et.al [19] analyzed the flows of AH in AC with/without particles by the theoretical and numerical methods, and found that their models can well predict the established and observed features that may exist in a traumatized eye. Heys et.al [20] presented a mathematical model of the anterior eye, and explored the elastohydrodynamic effects of accommodation on both the contour of the iris and the AH dynamics. In order to investigate heat transfer in the human eye, Karampatzakis & Samaras [21] presented a 3D numerical model, and found that consideration of AH fluid dynamics would affect the temperature distributions on the corneal and lens surfaces.

In conclusion, although great progress have been made in the study of AH dynamics in human eyes. However, there still are many unrevealed knowledge and challenges about the AH dynamics to be further explored [2]. At the same time, as far as we know, the studies on the AH dynamics are generally for a healthy eye [14-21], while few systematic studies on the AH dynamics are for a pathologic eye. The pathological changes of intraocular tissue may have an important effect on the AH dynamics, and some meaningful results may be obtained by studying them. In view of the methods, all above mentioned works have been completed either by theoretical methods or numerical ones. For the theoretical methods, as a series of approximations are necessary, the methods have limited applications in some special circumstances. As a potentially promising method, the numerical simulation has been successfully applied to various fields. However, as the inherent complex anatomical structure of the human eye, together with the high diversity of the tissues, the aqueous humor flow is a typical complex one, and to more clearly understand the AH dynamics is confronted with dire challenges. Therefore developing an effective model with simple algorithm, easily dealing with complex geometric boundaries and higher accuracies is significant.

Lattice Boltzmann method (LBM), as an alternative new computational hydrodynamics method, has achieved great progress in recent years [22-27]. The inherent advantages of simple algorithm, easy implementation of complex boundaries, high efficiency and full parallelization enable its successful applications on modeling complex fluid systems[28-32], including those in industry, science, bioengineering, biomedical engineering and so on.

In view of the dimension of human eye and the complex flow involving complex boundaries and coupling of the multi-physics problems, we can expect that LBM, a newly developed mesoscopic numerical method, may be a good candidate for modeling the dynamics of AH. However, to our best knowledge, there have been few reports of modeling the dynamics of AH with LBM. Thus it is quite natural that, this paper will focus on developing a coupled lattice Boltzmann model for the dynamics of AH in the human eye, and then checking its validity. By using the proposed model, we want to systematically investigate the roles of various macroscopic quantities, such as inlet flow rate, viscosity of AH, permeability of TM, anterior chamber angle and so on, in the dynamics of AH both for healthy and pathologic eyes, as well as their influences on IOP, and expect to obtain some interesting results.

The paper is organized as follows. First, in Sec. 2, we describe the development of a coupled lattice Boltzmann model for the dynamics of AH, including the macroscopic governing equations and the corresponding lattice Bhatnagar-Groose- Krook models. Sec. 3 starts to check the validity of the proposed model through a case of healthy eye, and then several additional cases are conducted to explore the effects of various parameters, such as inlet velocity, viscosity of AH, permeability of TM, the anterior chamber angle and so on, on IOP, and after that, the wall shear stresses on the corneal endothelium and the iris surface are investigated. Finally, we draw some conclusions.

## 2. Lattice Boltzmann model for the dynamics of AH

The dynamics of AH occurs in the anterior segment of eye, made up of cornea, TM, conjunctiva and sclera externally, and consisting of AC, iris, pupil, posterior chamber (PC) and ciliary body(CB) internally. It is characterized with behaviors of a multi-scale multi-physics coupled system.

### 2.1 Geometry

The anterior segment of eye consists of two chambers: the anterior and posterior chambers filled with AH. Based on the anatomy and physiological dimensions of the human eye [33], a 2D geometry resulting from a horizontal section is given in Fig.1, the posterior and anterior chambers of the eye is represented by $\Omega_1$. The section of the iris is expressed as $\Omega_2$, and the section of TM, an annular structure (in the basis of the cornea), is denoted as $\Omega_3$. The interface between the cornea and AC is represented by $\Gamma_1$, and the bottom surface between the lens and PC as $\Gamma_2$. All these sections and the interfaces are signaled in Fig.1 and the related geometric dimensions are listed in Tab. 1.

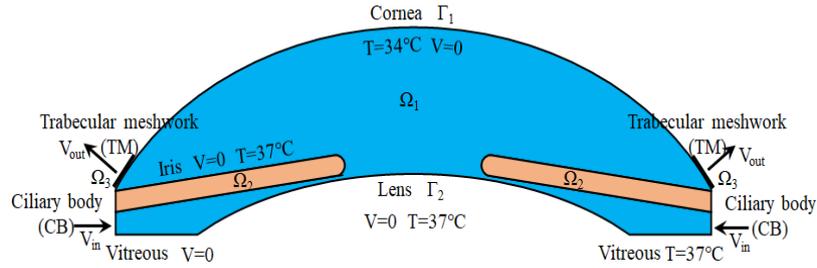

**Figure 1.** Geometry of the eye model.

**Table 1** Geometrical dimensions used in the model [15,20]

| Parameter | Value |
| --- | --- |
| Curvature radius of the cornea | 8.3mm |
| Distance between the cornea and lens along the vertical axis | 3.0mm |
| Thickness of the iris | 0.4mm |
| Anterior chamber angle | 45 ° |
| Pupil aperture | 3.0mm |
| Curvature radius of the lens | 12.5mm |
| Gap between the iris and lens | 0.05mm |
| Height of inlet (Ciliary Body) | 0.5mm |

## 2.2 Governing equations

To simulate the dynamics of AH in the anterior segment of eye, we firstly take into account that AH is a transparent and clear liquid with almost the same viscosity as saline water, can be treated as Newtonian. The cornea, iris and lens are assumed to be solid materials, without any changes in shape during AH flows, while TM is thought as an isotropic, homogenous, porous matrix.

### 2.2.1 Incompressible flow in the anterior and posterior chambers

Once aqueous humor is secreted from the ciliary body, at a flow rate of 2 to 2.5 μL/min, it flows forward to the posterior chamber, through the pupil into the anterior chamber, and then circulates there due to natural convection. In view of the very small flow rate and modest dimension, the flow of AH is creeping, compressibility and inertia can be neglected. Therefore, without considering the heat transfer effect, the flow of the AH in the posterior and anterior chambers can be described with the incompressible Navier-Stokes equations.

$$\nabla \cdot \boldsymbol{u} = 0 \,, \tag{1}$$

$$\frac{\partial \boldsymbol{u}}{\partial t} + \nabla \cdot (\boldsymbol{uu}) = -\nabla p + \nu \nabla^2 \boldsymbol{u} + \boldsymbol{F} \,, \tag{2}$$

where $\boldsymbol{u}$ is the velocity of AH, $\nu$ kinematic viscosity, $p$ the pressure in AH, and $\boldsymbol{F}$ is the body force. In order to solve Eqs.(1) and (2), no-slip boundary conditions are imposed on the surfaces $\Gamma_1$, $\Gamma_2$ and that of iris, while velocity boundary conditions are applied on AH inlet (at ciliary body) and outlet (at trabecular meshwork) boundaries.

## 2.2.2 Convection and diffusion of heat in anterior chamber

In this investigation, the effects from the convection and diffusion of heat, induced by the temperature difference cross AC, on the dynamics of AH are in consideration. In practice, there would be a temperature difference between the inner surface of the cornea, $\Gamma_1$, and that of the lens, $\Gamma_2$, thus the convection and diffusion of heat in AC is inevitable. It is known that if the viscous heat dissipation and compression work done by the pressure can be ignored, the temperature field is passively advected by the fluid flow and obeys a simple passive-scalar equation [34],

$$\frac{\partial T}{\partial t} + \nabla \cdot (\boldsymbol{u}T) = \theta \nabla^2 T, \tag{3}$$

where $T$ is the temperature in the temperature field, and $\theta$ the diffusivity. The heat flux boundary conditions with specific temperatures are imposed on the inner interfaces of the anterior and posterior chambers, which are coupled to Eq.(3).

## 2.2.3 Seepage of aqueous humor trough trabecular meshwork

TM is assumed as a cylindrical annular ring surrounding AC with specific thickness, approximatively described as an isotropic, homogeneous, porous matrix swollen with continuously flowing AH in this work. According to the porous media theory, the seepage of AH obeys the Darcy's law, which can be written as

$$\boldsymbol{u} = -\frac{\mathrm{K}}{\mu}\nabla p, \tag{4}$$

where $\boldsymbol{u}$ is the velocity of the seepage, $p$ the pressure in TM, $\mathrm{K}$ hydrodynamic permeability, and $\mu$ the viscosity of AH. The pressure boundary conditions with specific pressures are applied on the two sides of TM.

## 2.2.4 Coupling of multiple physics

In the present dynamics of AH, incompressible Navier-Stokes flows, convection and diffusion of heat and Darcy's flow are coexistence, which should be coupled with each other. As the flow of AH in the anterior and posterior chambers is natural convection, the well-known Boussinesq approximation can be used to couple the incompressible Navier-Stokes flow with the convection and diffusion of heat. In the Boussinesq approximation, the physical parameters of fluid such as density, viscosity and thermal diffusion coefficient are assumed to be constant except the density in external force term, where the fluid density $\rho$ is assumed to be a linear function of the temperature

$$\rho = \rho_0[1 - \beta(T - T_0)], \tag{5}$$

where $\rho_0$ and $T_0$ are the average fluid density and temperature, respectively, $\beta$ is the thermal expansion coefficient.

For the incompressible Navier-Stokes flow and the Darcy's flow, their deal is slightly simpler. At first IOP is calculated with Eq.(2), and then the velocity of the seepage can be obtained through the difference between IOP and the pressure of the vein, ultimately this desired velocity will be used as the velocity boundary condition, enabling the iteration of Eq.(2) to be repeated continually. With this algorithm, the coupling of Eq.(2) with Eq.(4) is implemented.

## 2.3 Lattice Boltzmann model for dynamics of AH

In lattice Boltzmann method, we use density distribution functions $f_i(\boldsymbol{x},t)$ to depict a fluid, and the distribution functions are represented the probability that a pseudo-fluid particle with velocity $\boldsymbol{e}_i$ comes into the lattice site $\boldsymbol{x}$ at time $t$. In order to mimic the motion of the pseudo-particles, at each time step, the pseudo-particles entering the same lattice site collide, and then the resulting distribution functions are streamed to the neighboring sites. The admissible velocities $\boldsymbol{e}_i$ ($i = 0, 1, \ldots z$), of component $e_{i\alpha}$, are dependent on the lattice topology, where z represents the lattice coordination number (i.e., the number of lattice links). Conventionally, $\boldsymbol{e}_0 = 0$ and $f_0(\boldsymbol{x},t)$ is the density distribution of particles at rest [35]. In single-relaxation-time approximation, the distribution functions obey the following evolvement equation in the lattice unit

$$f_i(\boldsymbol{x} + \boldsymbol{e}_i\delta t, t + \delta t) = f_i(\boldsymbol{x},t) - \frac{1}{\tau}\big[f_i(\boldsymbol{x},t) - f_i^{eq}(\boldsymbol{x},t)\big] + F_i, \qquad (6)$$

with $\delta x = c\delta t$ is the lattice spacing, $c$ the lattice velocity (magnitude), $\delta t$ the time step, and $F_i$ is a body force term if it exists. In the lattice Boltzmann dynamics of AH, it is obvious that the velocity field and the temperature field should be solved respectively by using two independent lattice Bhatnagar-Gross-Krook (BGK) equations, and then combined into a coupled system with the Boussinesq approximation.

As for the velocity field, the LBM for incompressible flow proposed by Guo [34] et.al is adopted, in which the equilibrium distribution function takes the form

$$f_i^{eq}(\boldsymbol{x},t) = \begin{cases} -4\sigma\dfrac{p}{c^2} + S_0(\boldsymbol{u}) & i = 0 \\ \lambda\dfrac{p}{c^2} + S_i(\boldsymbol{u}) & i = 1,2,3,4 \\ \gamma\dfrac{p}{c^2} + S_i(\boldsymbol{u}) & i = 5,6,7,8 \end{cases}, \qquad (7)$$

where $\sigma$, $\lambda$ and $\gamma$ are parameters satisfying $\lambda + \gamma = \sigma$ and $\lambda + 2\gamma = 1/2$. In Eq. (7), $S_i(\boldsymbol{u})$ is a function of macroscopic velocity $\boldsymbol{u}$ and discrete velocity $\boldsymbol{e}_i$

$$S_i(\boldsymbol{u}) = \omega_i\left[3\frac{(\boldsymbol{e}_i \cdot \boldsymbol{u})}{c} + \frac{9}{2c^2}(\boldsymbol{e}_i \cdot \boldsymbol{u})^2 - \frac{3}{2c^2}|\boldsymbol{u}|^2\right], \quad i = 0,1,2,\ldots,8, \qquad (8)$$

with the weight coefficients $\omega_0 = 4/9$, $\omega_1 = \omega_2 = \omega_3 = \omega_4 = 1/9$ and $\omega_5 = \omega_6 = \omega_7 = \omega_8 = 1/36$ for D2Q9 topology. The primitive macroscopic variables of the incompressible fluid, the velocity $\boldsymbol{u}$ and pressure $p$, are given by

$$\boldsymbol{u} = \sum_{i=0}^{8} c\,\boldsymbol{e}_i f_i, \qquad p = \frac{c^2}{4\sigma}\left[\sum_{i=1}^{8} f_i + S_0(\boldsymbol{u})\right]. \qquad (9)$$

Through a multi-scaling expansion procedure, the incompressible Navier-Stokes equations (1)

and (2) can be recovered from this incompressible lattice BGK model to the order of $O(\delta t^2)$ or $O(\delta x^2)$, if the microscopic velocity $c = O(1)$, in which the kinematic viscosity $\nu$ is given by

$$\nu = \frac{2\tau - 1}{6}\frac{(\delta x)^2}{\delta t}. \tag{10}$$

The corresponding no-slip boundary conditions on the surfaces of $\Gamma_1$, $\Gamma_2$ and that of iris are implemented with the method proposed by Bouzidi et.al [36], and the velocity boundary conditions, on the inlet (ciliary body) and the outlet (trabecular meshwork), are evaluated by the algorithm presented by Guo et.al [37].

In regard to the temperature field, a lattice with four discrete velocity directions $\boldsymbol{e}_1$, $\boldsymbol{e}_2$, $\boldsymbol{e}_3$ and $\boldsymbol{e}_4$ is introduced and thus the lattice BGK equation for Eq.(3) is written as [34]

$$T_i\big(\boldsymbol{x} + c\boldsymbol{e}_i\,\delta t, t + \delta t = T_i(\boldsymbol{x},t)\big) - \frac{1}{\tau_T}\big[T_i(\boldsymbol{x},t) - T_i^{eq}(\boldsymbol{x},t)\big], \tag{11}$$

where $\tau_T$ is the dimensionless relaxation time, $T_i$ is the temperature distribution function, and $T_i^{eq}$ is the equilibrium distribution function given by

$$T_i^{eq} = \frac{T}{4}\Big[1 + 2\frac{\boldsymbol{e}_i \cdot \boldsymbol{u}}{c}\Big]. \tag{12}$$

The fluid temperature $T$ can be calculated from the temperature distribution function

$$T = \sum_{i=1}^{4} T_i(\boldsymbol{x},t). \tag{13}$$

Similar to the recovering procedures of Eq.(2), through a multi-scaling expansion, the macroscopic Eq.(3) can be recovered from Eq.(11) to the $O(\delta t^2)$ order if Mach number, M, is of the same order of $\delta t$ or higher, where the diffusivity $\theta$ is determined by

$$\theta = \frac{2\tau_T - 1}{4}\frac{(\delta x)^2}{\delta t}. \tag{14}$$

In the evolution of Eq.(11), the thermodynamic boundary conditions with specific temperatures on the inner surfaces of the AC and PC are evaluated with the extrapolation method [34].

In order to coupling Eq.(6) with Eq.(11) in the lattice Boltzmann simulation, the Boussinesq approximation has to be performed in LBM. In our simulations, the body force is just the buoyant force, and thus the density of the body force, $\boldsymbol{F}$, in Eq.(2) can be explicitly written as

$$\boldsymbol{F} = \boldsymbol{g} - \boldsymbol{g}\beta(T - T_0), \tag{15}$$

where $\boldsymbol{g}$ is the acceleration vector of gravity. Correspondingly, after absorbing the first constant part of $\boldsymbol{F}$ into the pressure term, the macroscopic Eq.(2) finally becomes

$$\frac{\partial \boldsymbol{u}}{\partial t} + \nabla \cdot (\boldsymbol{uu}) = -\nabla p + \nu\nabla^2 \boldsymbol{u} - \boldsymbol{g}\beta(T - T_0). \tag{16}$$

Consequently, the coupling of Eqs.(2) and (3), through the Boussinesq approximation, comes down to discretizing the density of the body force, $\boldsymbol{F}$, in Eq.(15) into the force term, $F_i$, in Eq.(6). By using the forcing technology proposed in Ref.38, this discretization can be performed as

$$F_i = -\frac{\delta t}{2c}\alpha_i \boldsymbol{e}_i \cdot \boldsymbol{g}\beta(T - T_0), \tag{17}$$

where $i=2$ and $i=4$ refer to the direction of gravity, and when $i=2$ and 4, $\alpha_i = 1$, otherwise $\alpha_i = 0$.

With respect to the seepage flow through TM, a relative simpler averaging method is used for solving this flow. TM is assumed as a porous media with a finite thickness $h_{TM}$, which locates between AC and SC. We know that the pressure in AC is just IOP, $p_{IOP}$, and that the observed pressure in SC, $p_0$, equals to that in blood. If the pressure in TM, $p_{TM}$ is supposed to vary linearly along the thickness direction, its gradient will be a constant and equal to the pressure difference cross TM divided by the thickness $h_{TM}$, e.g. $\nabla p_{TM} = \frac{p_{IOP} - p_0}{h_{TM}}$. In this mentality, the coupling procedure of the incompressible Navier-Stokes flow and Darcy's flow can be conducted as follows: firstly, we can evaluate IOP from evolving Eq.(6); secondly, the seepage velocity can be derived from the pressure gradient in TM, $\nabla p_{TM}$; and finally the calculated seepage velocity is applied to the velocity boundary condition on AC. At this end, the coupling of the velocity field and the seepage flow through TM has been carried out.

## 3. Lattice Boltzmann dynamics of AH

In order to well understand in vivo dynamics of AH, the flow patterns and influences of various factors, such as AH secretion rate, TM permeability and so on, should be investigated. For this matter, very few clinical and laboratorial results are available and numerical simulations may be significant methods to address these problems. Toward this end, several cases of the AH flow in both healthy (with the TM permeability given by K=7$\times 10^{-15}$ m$^2$) and pathological (with the TM permeability given by K=2.3$\times 10^{-15}$m$^2$) [6] situations are simulated with the present model.

### 3.1 Flow of AH in healthy eye

Due to insufficient laboratorial results, the direct verification of the proposed model may encounter with difficulties, and thus an indirect validation with now available experimental data would be performed. It is known that the available experimental data for healthy eyes are much more plentiful compared to that for pathologic ones. Therefore we will begin with the simulations for the flow of AH in healthy eye, which will be served as the basis for further researches. Let's consider a flow of AH in an anterior segment of healthy eye in standing orientation, with the geometrical dimensions listed in Tab. 1 and the physical parameters in Tab.2.

**Table 2** Physical parameters used in the model [39]

| Parameter | Value |
| --- | --- |
| AH viscosity, μ | 7.1$\times 10^{-4}$ Pa s |
| AH density, ρ | 9.9$\times 10^{2}$ kg/m$^3$ |
| AH thermal conductivity, k | 0.58 W/(mK) |
| AH specific heat, Cp | 4.2$\times 10^{3}$ J/(kgK) |
| Volume expansion coefficient, β | 3.2$\times 10^{-4}$ ℃$^{-1}$ |
| Cornea surface temperature, T$_{ref}$ | 34 ℃ |

The simulations are performed on the domain with dimensions 588 ×1921 (in lattice units). The normal production rate of AH is taken to be 2.5μL/min, the corresponding inlet velocity is $2 \times 10^{-6}$m/s, the temperature on surface $\Gamma_2$ is equal to 37 ℃, which is the core body temperature. In healthy eye, the normal IOP is usually 1950 Pa, and the pressure observed in SC is 1200 Pa, which is the same as that in the blood. As for the temperature, the normal temperature difference across AC is usually 3 ℃ [39]. In the flow of AH, it is well known that, if the production rate is higher than the outflow rate, AH will accumulate in AC and PC, which should lead to an increase of IOP; conversely, if the production rate is lower than the outflow rate, AH in AC and PC will decrease, and IOP would decline. Therefore the flow of AH will reach equilibrium with constant IOP if and only if the production rate is equal to the outflow rate. In our model of healthy eye, all geometrical and physical parameters are set to be the values within their normal ranges except for the thickness of TM, which will be used as an adjustable parameter for controlling the outflow rate (see Eq.(4)). Initially, IOP is set to be a deviated value from its normal one, such as 1948 Pa or 1952 Pa, while the thickness of TM is set to be a testing value. By adjusting this thickness to a proper value, the stable flow of AH with constant IOP can be always achieved. In term of this method, we perform the simulations through carefully adjusting the thickness, which will be taken as the effective thickness of TM, and finally obtain the stable flow of AH with constant IOP of 1950 Pa. The simulated results for IOP and the outflow rate are drawn in Fig.2, which indicates that, with the time increase, IOP and the outflow rate gradually converge to 1950 Pa and inflow rate respectively. This illustrates that, due to the negative feedback of IOP, the model can always reach the equilibrium flow of AH by itself, and therefore it is convergent and robust. By changing other parameters, the simulations may reach other equilibrium flow states. Thus the proposed model for healthy eye can be applied as the basis of the further researches on pathological eyes.

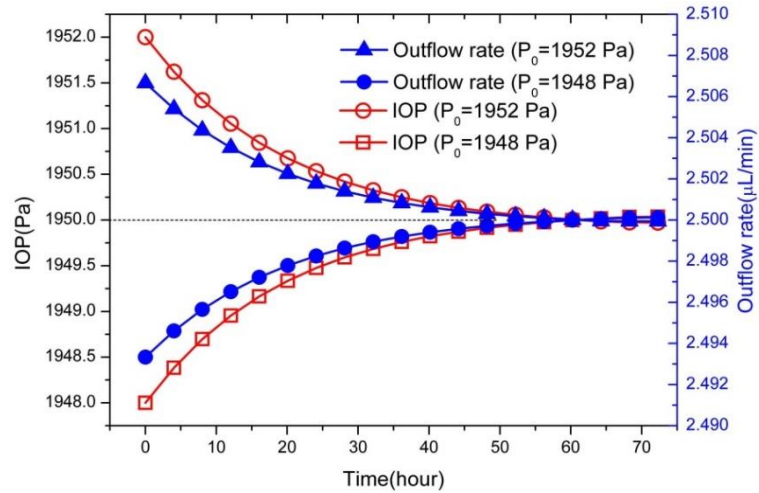

**Figure 2.** IOP and outflow rate vs time, initial pressure $P_0$=1948, 1952 (Pa) respectively.

Thermally driven natural convection plays an important role on the dynamics of AH in the human eye [21]. In order to understand the importance of this heat effect, we also perform the simulations without the temperature gradient across AC as a comparison, and the simulated results for the flow of AH with and without temperature gradient are given in Figs.3 and 4. Fig.3 (left) shows that, without temperature gradient, the flow of AH is fully generated by the AH production rate, and is rather weak with the maximum velocity of $3.03 \times 10^{-5}$ m/s and no vortex formation, which could affect the normal physiological functions of AH. With the temperature difference of 3 ℃ across AC (Fig.3 (right)), it is clearly that the flow of AH becomes obviously strong with the maximum velocity of $7.05 \times 10^{-4}$ m/s, in which buoyancy is the dominant driving mechanism, and the maximum velocity is in good agreement with the one calculated by Heys and Barocas ($7 \times 10^{-4}$ m/s) [39].

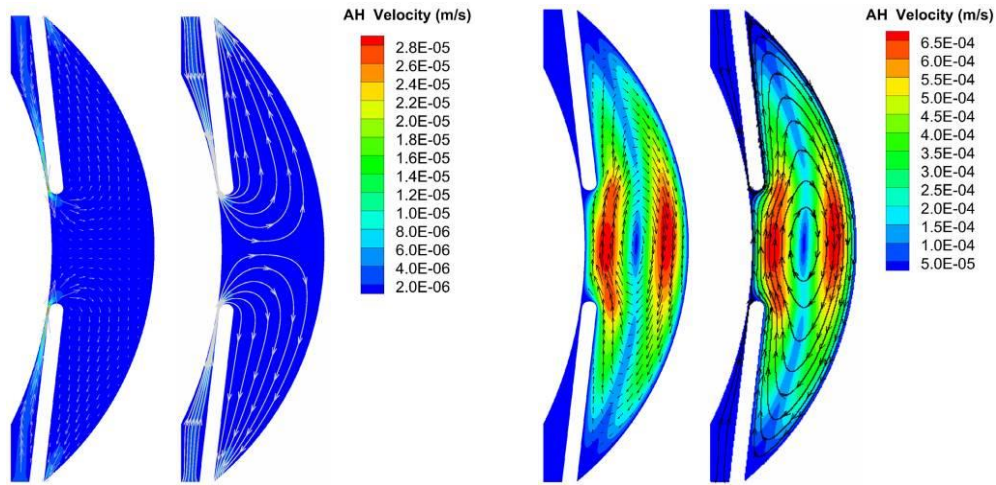

**Figure 3.** Velocity vectors and stream lines of the AH flow with (right) and without (left) temperature gradient for the healthy eye in standing orientation.

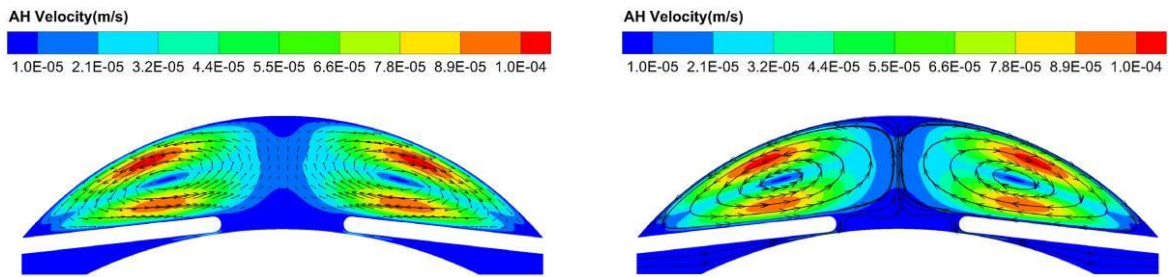

**Figure 4.** Velocity vectors and stream lines of the AH flow with temperature difference of 3 ℃ for healthy eye at up-facing position.

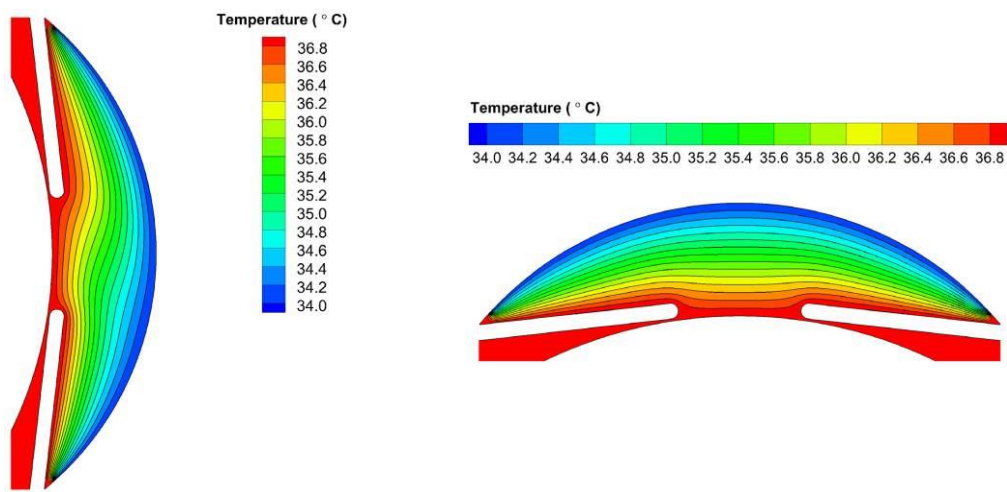

**Figure 5.** Temperature contours in AC with temperature difference of 3 ℃ in both standing and up-facing eyes.

Furthermore, a strong recirculation vortex with its center near the center of AC is formed. For further investigating the recirculation of AH in AC, which is beneficial to understand the formation of Krukenberg's Spindle, we simulate the flow of AH in both standing and up-facing eyes with different temperature gradients $\Delta T$, and the results are listed in Tab.3 and drawn in Figs.6 and 7. For vortex flow, the vortex intensity is defined as

$$I = 2\int_\Gamma \omega_z \, ds, \qquad \omega_z = \frac{1}{2}(\frac{\partial u_y}{\partial x} - \frac{\partial u_x}{\partial y}) \ , \qquad (18)$$

where $\omega_z$ is the angular velocity of AH, $\Gamma$ is the area of AC, $u_x$ and $u_y$ are the two velocity components of AH respectively.

**Table 3** The maximum velocities, vortex intensities and positions of the vortex center (standing)

| Temperature of cornea (℃) | Temperature Difference (℃) | Maximum velocity (m/s) | Vortex intensity | Position of vortex center (mm) |
|---|---|---|---|---|
| 32 | 5 | $1.09\times10^{-3}$ | 0.469832 | (1.871,4.846) |
| 33 | 4 | $9.05\times10^{-4}$ | 0.386765 | (1.872,4.841) |
| 34 | 3 | $7.05\times10^{-4}$ | 0.298784 | (1.872,4.834) |
| 35 | 2 | $4.88\times10^{-4}$ | 0.204950 | (1.872,4.824) |
| 36 | 1 | $2.52\times10^{-4}$ | 0.104749 | (1.872,4.811) |
| 36.8 | 0.2 | $5.09\times10^{-5}$ | 0.020917 | (1.872,4.788) |

From Tab.3 and Fig.6 (in the standing eye), we find that, with the increase of the temperature gradient, the maximum velocity and vortex intensity increase, and its center position rises slightly along the vertical direction. However, for the cases in the up-facing eye with temperature difference of 3℃ across AC, Fig.4 shows that, driven by the temperature gradient, different from that in the standing eye, two symmetric vortices appear in AC with equal maximum velocity of $1.05\times10^{-4}$ m/s, about 7 times smaller than that in the standing eye. Fig.7 tells us that, just like that in the standing eye, with the increase of the temperature gradient, the maximum velocity and vortex intensity also increase, and the two centers of the vortices rise along the vertical direction as well, but accompanying with their moving toward the center of AC simultaneously. Fig.5 gives the distributions of temperature in both standing and up-facing eyes with temperature difference of 3℃ across AC. Fig.5 (left) shows that, in the standing eye, the natural convection currents cause the posterior cornea temperature to vary, and the symmetry about the central horizontal axis of the temperature distribution is broken. Fig.5 (right) indicates that, in up-facing eye, although the natural convection currents also cause the posterior cornea temperature to vary, the temperature distribution is symmetrical to the mid-perpendicular of AC due to the symmetry of the two vortices.

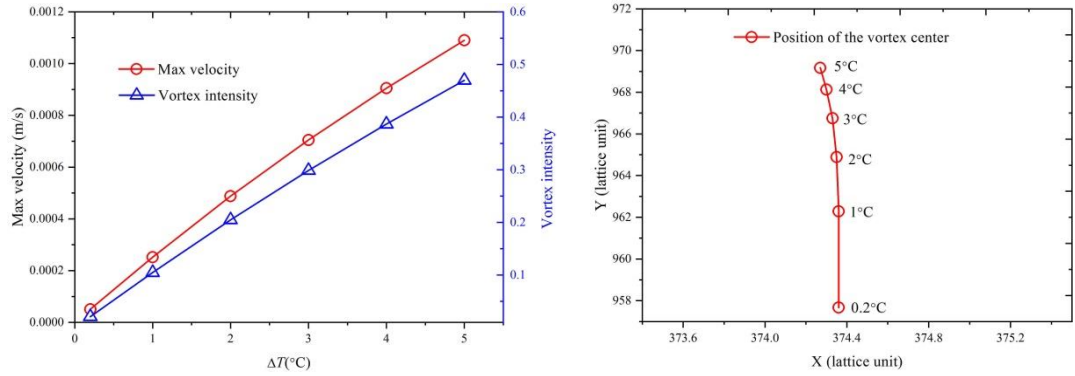

**Figure 6.** The maximum velocities, vortex intensities (left) and positions of the vortex center (right) in the standing eye with different temperature difference $\Delta T$.

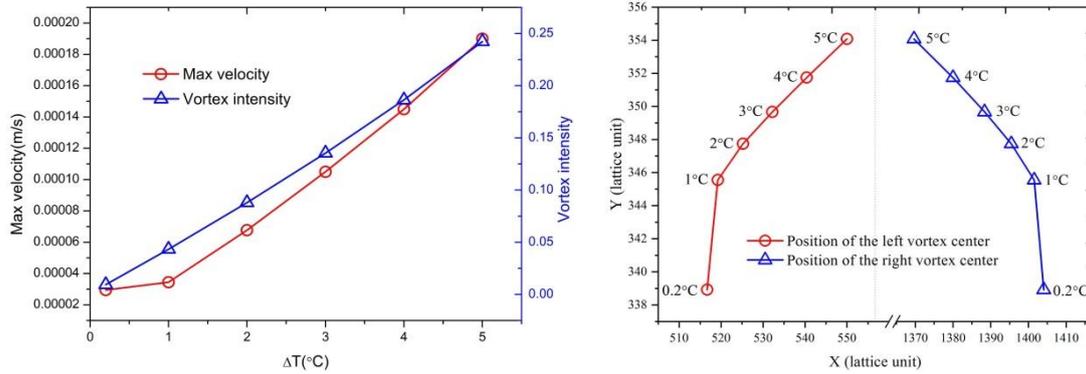

**Figure 7.** The maximum velocities, vortex intensities (left) and positions of the vortex center (right) in the up-facing eye with different temperature difference $\Delta T$.

## 3.2 Influence of aqueous humor secretion rate on the flow of AH

It is well-know that IOP can be controlled by decreasing the production rate of AH. Thus the influence of AH secretion rate on IOP or the flow of AH is significant. On the basis of the simulations for healthy eye, simulations for the flow of AH with changing secretion rates are performed. With an altered secretion rate of AH, the simulation will converge to an equilibrium state with a new IOP or outflow rate. In order to understand the details of this influence on the flow of AH in both healthy (with the permeability of $K=7\times10^{-15}m^2$) and pathologic (with the permeability of $K=2.3\times10^{-15}m^2$) [6] eyes, we investigate the two corresponding cases with changing secretion rates. The simulated results are shown in Tab.4 and Fig.8. For comparisons, we also list the results obtained by Ref.6 in Tab.4, in which the relative error is defined as $\frac{V_s - V_{ref}}{V_{ref}} \times 100\%$, where $V_{ref}$ and $V_s$ are the values by Ref.6 and the present model respectively.

**Table 4** IOP in healthy and pathologic eyes for different values of inflow velocity

| Velocity(m/s) | Pressure in a pathologic eye (Pa) | | Relative error | Pressure in a healthy eye (Pa) | | Relative error |
|---|---|---|---|---|---|---|
| | **Ref.6** | **present model** | | **Ref.6** | **present model** | |
| 0.5$V_{in}$ | 2548 | 2327 | -8.67% | 1585 | 1578 | -0.44% |
| 1.0V$_{in}$ | 3896 | 3540 | -9.14% | 1970 | 1950 | -1.01% |
| 1.25V$_{in}$ | 4570 | 4085 | -10.61% | 2163 | 2147 | -0.74% |
| 1.5V$_{in}$ | 5244 | 4683 | -10.70% | 2335 | 2332 | -0.13% |
| 1.75V$_{in}$ | 5917 | 5222 | -11.75% | 2548 | 2525 | -0.90% |
| 2.0V$_{in}$ | 6591 | 5838 | -11.42% | 2741 | 2717 | -0.88% |

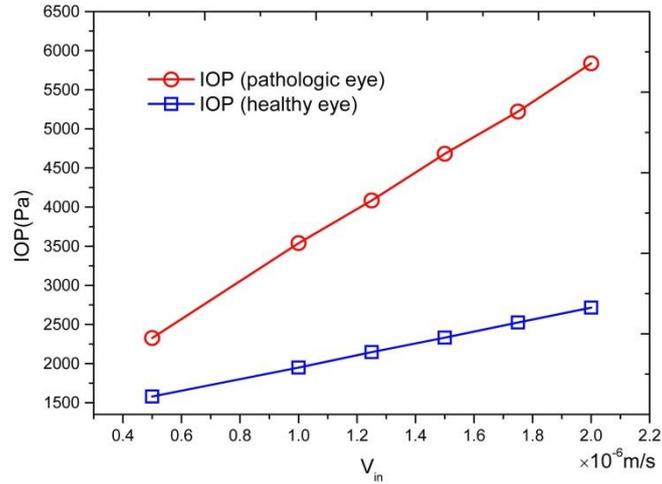

**Figure 8.** IOP in AC as a function of the AH production rate

From Tab.4 and Fig.8, we see that our simulated results are very close to those obtained in Ref.6, and with decrease of the inlet velocity, IOP linearly reduces in both healthy and pathologic situations. Furthermore, the velocity of IOP reduction is quit faster in pathologic eye than in healthy eye. This numerically illustrates that IOP can be effectively lowered by decreasing the AH secretion rate with drugs, especially in pathologic situations.

### 3.3 Influence of resistance of aqueous humor drainage on the flow of AH

We also know that IOP can be controlled by reducing the level of resistance of AH drainage. On the basis of the simulations for healthy eyes, simulations for the flow of AH, with changing permeability of TM and keeping the other parameters unchanged, are carried out. When the permeability of TM has changed, the simulation will reach a new equilibrium state with the corresponding IOP or outflow rate. For comparing with the results obtained in Ref.6, the simulations with the same permeability or porosity values as those used in Ref.6 are performed. The simulated results with the present model, together with those from Ref.6, are listed in Tab.5 and drawn in Fig.9 simultaneously.

**Table 5** IOP for different grades of obstruction of the TM.

| Porosity (ε) | Permeability (m$^2$) of TM | IOP (Pa) in | | Relative error |
|---|---|---|---|---|
| | | Ref.6 | present model | |
| 0.4 | $7.59 \times 10^{-14}$ | 1271 | 1270 | -0.11% |
| 0.3 | $2.35 \times 10^{-14}$ | 1429 | 1426 | -0.24% |
| 0.25 | $1.19 \times 10^{-14}$ | 1655 | 1647 | -0.51% |

| 0.225 | $8.09 \times 10^{-15}$ | 1867 | 1856 | -0.62% |
| 0.2 | $5.33 \times 10^{-15}$ | 2211 | 2194 | -0.78% |
| 0.175 | $3.36 \times 10^{-15}$ | 2805 | 2779 | -0.94% |
| 0.15 | $1.99 \times 10^{-15}$ | 3905 | 3870 | -0.90% |
| 0.125 | $1.09 \times 10^{-15}$ | 6154 | 6065 | -1.45% |
| 0.1 | $5.27 \times 10^{-16}$ | 11437 | 11254 | -1.60% |

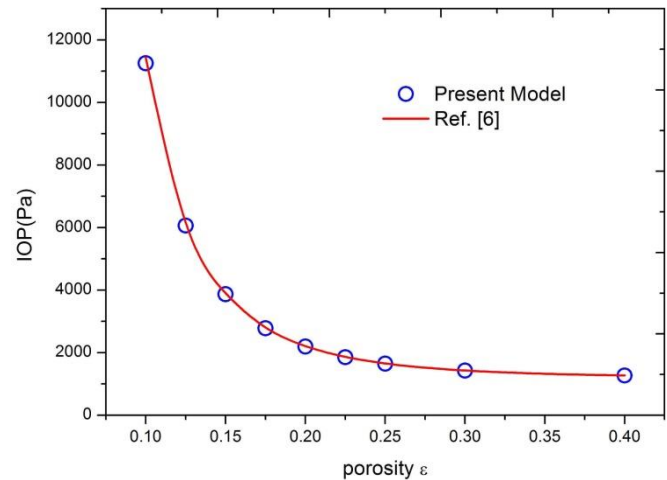

**Figure 9.** Dependence of IOP on the porosity ε.

From Tab.5 and Fig.9, we see that our simulated results are well agreement with those obtained in Ref.6, and with the increase of the porosity (or permeability) of TM, IOP reduces rapidly. More specifically, the velocity of IOP reduction is non-uniform, in which IOP suffers a slight increase when the permeability K is in the interval of $[7.59 \times 10^{-14}, 5.33 \times 10^{-15}]$ (m$^2$) and but exhibits a steep gradient when K is in the interval of $[3.36 \times 10^{-15}, 5.27 \times 10^{-16}]$ (m$^2$). This numerically illustrates that IOP can be efficiently lowered by increasing the permeability (or porosity) of TM, especially when the permeability is below the value of $3.36 \times 10^{-15}$ m$^2$.

### 3.4 Influence of the viscosity of aqueous humor on the flow of AH

Viscosity of AH is a basic physical quantity associated to transport of momentum among the layers of AH, which plays an important role in the flow of AH. However, the influences of AH viscosity on the flow of AH have not been fully addressed so far. Therefore, the investigations on the influence of viscosity on the flow of AH are still needed. On the basis of the simulations for healthy eyes, simulations for the flow of AH with different viscosities are carried out. When the viscosity has changed, the simulation will eventually reach an equilibrium state with a new IOP or outflow rate. We investigate two cases of both healthy and pathologic eyes. The simulated results are shown in Tab.6 and Fig.10.

**Table 6** IOP dependence on fluid viscosity $\mu$

| Fluid viscosity $\mu$ (Pa s) | IOP (Pa) in a healthy eye | IOP (Pa) in a pathologic eye |
| --- | --- | --- |
| $6.947 \times 10^{-4}$ | 1950 | 3540 |
| $7.225 \times 10^{-4}$ | 1985 | 3600 |
| $7.523 \times 10^{-4}$ | 2017 | 3698 |
| $7.840 \times 10^{-4}$ | 2061 | 3805 |
| $8.180 \times 10^{-4}$ | 2094 | 3917 |
| $8.545 \times 10^{-4}$ | 2135 | 4038 |

From Tab.6 and Fig.10, we see that, with the increase of the viscosity, IOPs in both healthy and pathologic eyes will all rise at different velocities, and the IOPs rise faster in pathologic eyes than in the healthy ones. But, on the other hand, the IOP increase is insensible to the viscosity of AH, compared to the permeability of TM and AH secretion rate.

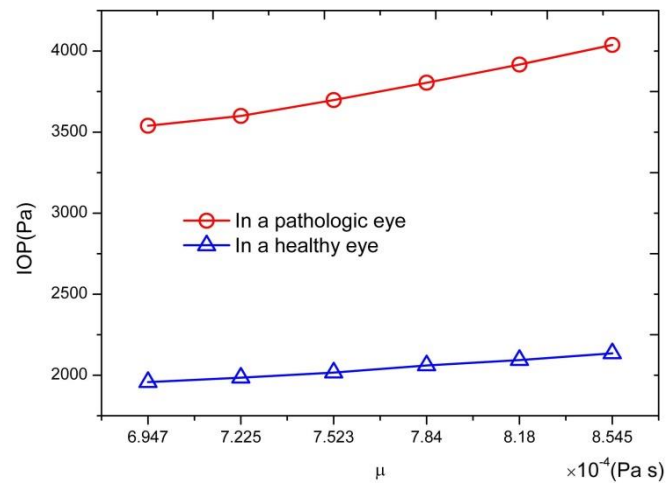

**Figure 10.** Dependence of IOP on the fluid viscosity $\mu$.

## 3.5 Influence of anterior chamber angle on the flow of AH

Anterior chamber angle is defined as the attachment angle of the iris to the cornea, which is closely related with close-angle glaucoma. In order to research the IOP dependence on anterior chamber angle, we perform the simulations with different anterior chamber angles in both healthy and pathologic eyes, in which the range of the angle is much larger than that in the Ref.15. The simulated results are listed in Tab.7 and drawn in Fig.11. Both Tab.7 and Fig.11 indicate that, with the decrease of the anterior chamber angle, the IOPs increase in both healthy and pathologic eyes with increasing velocities, of which the IOP increase velocity is slightly greater in the pathology eye than in the healthy one. Analyzing in detail, we find again that IOP increases slowly with the decrease of the anterior chamber angle when this angle is above some critical value (about 15 ℃), and it dramatically increases once the angle is lower than the critical value, which implies a higher risk of glaucoma.

**Table 7** IOP dependence on anterior chamber angle

| Anterior chamber angle(°) | IOP in a healthy eye(Pa) | IOP in a pathologic eye(Pa) |
|:---:|:---:|:---:|
| 10 | 2006 | 3725 |
| 15 | 1980 | 3640 |
| 20 | 1974 | 3617 |
| 25 | 1968 | 3598 |
| 30 | 1965 | 3586 |
| 35 | 1955 | 3562 |
| 40 | 1952 | 3548 |
| 45 | 1950 | 3540 |

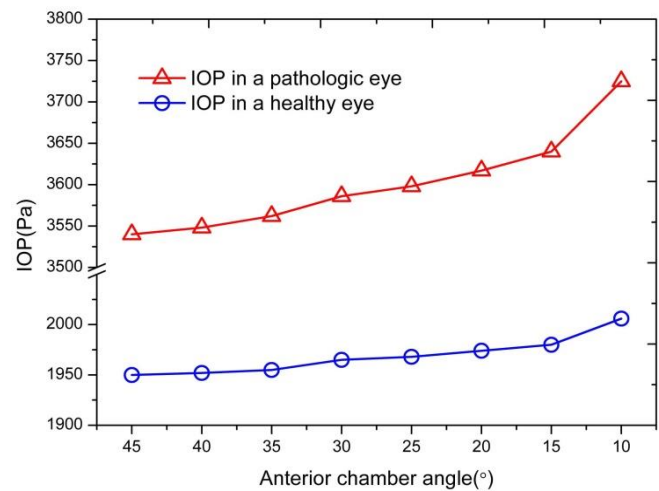

**Figure 11.** IOP dependence on anterior chamber angle

### 3.6 Wall shear stress

The flow of AH in the anterior chamber will form a wall shear stress (WSS) on the corneal endothelium or the iris surface. Researches show that WSS can change the endothelial cells' functions and structures [40,41]. In this subsection, the factors that may affect these wall shear stresses are investigated in a healthy eye. The shear stress contour for standing and up-facing orientations of the eye are plotted in Fig. 12 with a temperature difference across AC $\Delta T$=3 ℃. It can be seen that there is an obvious difference for the distribution of WSS on the corneal endothelium and the iris surface between the standing and the up-facing orientation of the eye. For the standing orientation, the maximum cornea WSS is located at the top of cornea and decreases towards the anterior chamber angle, while the maximum iris WSS is located in the inner periphery of the iris close to the pupil and lessens towards the anterior chamber angle. For the up-facing orientation, the maximum cornea WSS is located in the midperiphery of the cornea and reduces towards the anterior chamber angle and the top of cornea respectively, and the maximum iris WSS appears in the middle of the iris surface and decreases towards the pupil and the anterior chamber angle separately.

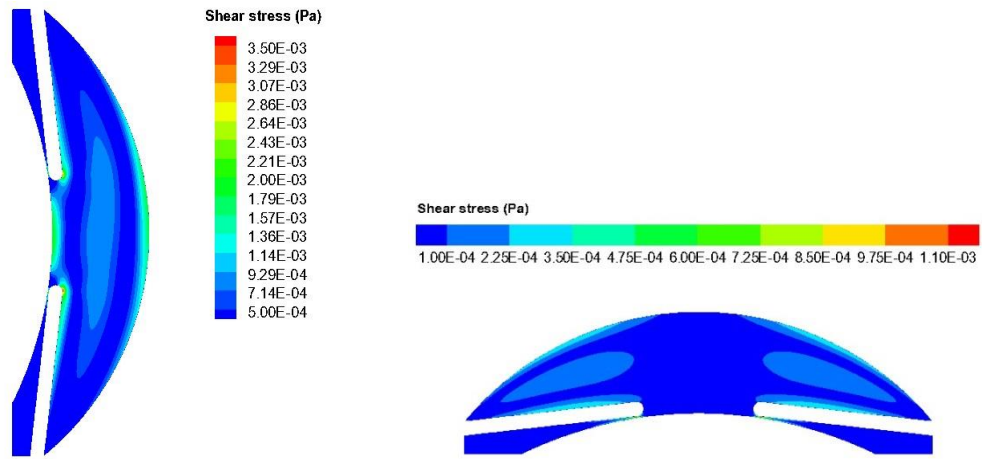

**Figure 12.** Shear stress contour for the anterior chamber. Standing orientation (left) and up-facing orientation (right) with temperature difference $\Delta T$=3 ℃.

**Table 8** Average cornea and iris WSSes variation with temperature difference $\Delta T$

| Tissue | Eye-orientation | WSS with different $\Delta T$ (Pa) | | | |
|---|---|---|---|---|---|
| | | **0.2 ℃** | **3 ℃** | **5 ℃** | **7 ℃** |
| Cornea | Standing | $5.76 \times 10^{-5}$ | $8.36 \times 10^{-4}$ | $1.36 \times 10^{-3}$ | $1.85 \times 10^{-3}$ |
| | Up-facing | $1.22 \times 10^{-5}$ | $1.66 \times 10^{-4}$ | $2.85 \times 10^{-4}$ | $5.48 \times 10^{-4}$ |
| Iris | Standing | $6.57 \times 10^{-5}$ | $7.73 \times 10^{-4}$ | $1.21 \times 10^{-3}$ | $1.61 \times 10^{-3}$ |
| | Up-facing | $1.76 \times 10^{-5}$ | $1.85 \times 10^{-4}$ | $3.24 \times 10^{-4}$ | $5.67 \times 10^{-4}$ |

**Table 9** Average cornea and iris WSSes variation with inflow velocity, permeability of TM and AH viscosity

| Parameter | Value | Wss in standing orientation (Pa) | | Wss in up-facing orientation (Pa) | |
|---|---|---|---|---|---|
| | | **Cornea** | **Iris** | **Cornea** | **Iris** |
| Inflow velocity | $1.00 \times 10^{-6}$ | $8.360 \times 10^{-4}$ | $7.668 \times 10^{-4}$ | $1.653 \times 10^{-4}$ | $1.828 \times 10^{-4}$ |
| (m/s) | $4.00 \times 10^{-6}$ | $8.364 \times 10^{-4}$ | $7.836 \times 10^{-4}$ | $1.682 \times 10^{-4}$ | $1.904 \times 10^{-4}$ |
| | | 0.05% | 2.19% | 1.75% | 4.16% |
| Permeability of | $5.27 \times 10^{-16}$ | $8.361 \times 10^{-4}$ | $7.737 \times 10^{-4}$ | $1.663 \times 10^{-4}$ | $1.852 \times 10^{-4}$ |
| TM (m$^2$) | $7.59 \times 10^{-14}$ | $8.361 \times 10^{-4}$ | $7.737 \times 10^{-4}$ | $1.663 \times 10^{-4}$ | $1.852 \times 10^{-4}$ |
| | | 0.00% | 0.00% | 0.00% | 0.00% |
| AH viscosity (Pas) | $6.947 \times 10^{-4}$ | $8.359 \times 10^{-4}$ | $7.730 \times 10^{-4}$ | $1.663 \times 10^{-4}$ | $1.852 \times 10^{-4}$ |
| | $8.545 \times 10^{-4}$ | $8.462 \times 10^{-4}$ | $8.015 \times 10^{-4}$ | $1.644 \times 10^{-4}$ | $1.826 \times 10^{-4}$ |
| | | 1.23% | 3.69% | -1.14% | -1.40% |

Tab. 8 shows the dependence of average cornea and iris WSSes with temperature difference $\Delta T$. It can be seen that there is a linear relationship between the WSSes and the temperature difference. The greater the temperature difference, the greater WSS is. The WSS is sensitive to the temperature difference, when the temperature difference increases from 0.2 ℃ to 7 ℃, there is an increasing more than an order of magnitude on the cornea or iris WSS in both standing and up-facing orientations of the eye. For instance, the cornea WSS increased by about 45 times in the up-facing orientation. Tab. 9 shows the dependence of average cornea and iris WSSes with inflow velocity (AH secretion rate), permeability of TM and AH viscosity. For the inflow velocity, the average WSSes in standing and up-facing orientation increase by 0.05% and 1.75% for cornea, 2.19% and 4.16% for iris respectively when it increases by 4 times. The increase rate of the average WSS in up-facing orientation is slightly higher than that in standing orientation. For the permeability of TM, according to Eq.(4), it is directly proportional to outflow velocity of AH. However, we found that the average cornea and iris WSSes remained almost unchanged when the permeability of TM increases by144 times. The cause may be that the gradient of AH velocity near the wall of the cornea and iris do not increase although the outflow velocity of AH increases. In order to investigate the effect of the AH viscosity on WSS, we increased the AH viscosity by 1.23 times, The average cornea and iris WSSes in standing orientation increase by 1.23% and 3.69%, while those in up-facing orientation decrease by -1.14% and - 1.40%, respectively. According to Newton's law of viscosity, the viscosity and velocity gradient of AH are in direct proportion to the shear stress. However, the velocity gradient of AH near the wall decreases when the viscosity of AH increases, so the average cornea and iris WSSes in standing orientation have a slight increase. On the contrary, the WSSes in up-facing orientation have a little reduce due to the influence of the velocity gradient decreasing on the WSS is greater than that of the AH viscosity. The same simulations mentioned above were also performed in a pathologic eye. However, we found that the results are almost the same as those in the healthy eye. The cause is that the healthy and the pathologic eyes are defined by the permeability of TM $K=7\times10^{-15}m^2$ and $K=2.3\times10^{-15}m^2$ respectively in this paper, and the above results have proved that the permeability of TM has little effect on the WSSes. Through analyzing above simulation results, we can draw conclusions that the average cornea and iris WSSes in standing orientation is significantly higher than those in up-facing orientation, and among the factors affecting the WSS, the temperature difference has the greatest influence on the WSSes, while the influence of inflow velocity, permeability of TM and AH viscosity are not obvious.

The magnitude of the WSSes on the iris surface and corneal endothelium is an important parameter for the prevision of endothelial cell detachment in pigment dispersion syndrome, pigmentary glaucoma and bullous keratopathy [42,43]. In our simulations, the average the maximum cornea and iris WSSes are $1.61\times10^{-3}$Pa and $1.85\times10^{-3}$Pa respectively. According to Gerlach et al. [44], the magnitudes required to detach endothelial cell on iris surface are in the rage between 0.51 and 1.53 Pa, while the ones required to detach endothelial cell on corneal endothelium are comprised between 0.01 and 1.0 Pa as reported by Kaji et al. [45]. Due to the WSSes in our cases are much lower than the detaching threshold value for such cells, it is unlikely that the WSS could detach pigments from the anterior iris surface or endothelial cells from the corneal endothelium, but the plots of shear stress give insight about the location and probability of pigments detachment on the iris surface or corneal endothelial cells loss on the corneal endothelium  in case of some specific disease，such as angle-closure glaucoma after laser iridotomy.

## 4. Conclusion

A coupled lattice Boltzmann model for simulating the dynamics of aqueous humor in the human eye is presented, which involves incompressible Navier-Stokes flow, heat convection and diffusion, and Darcy seepage flow. The verification simulations for the healthy eye show that the present model is available, convergent and robust due to its pressure negative feedback mechanism. Further researches on the dynamics of AH in the healthy eyes indicate that heat convection and diffusion is crucial to the normal physiological functions of AH. Furthermore, the flow patterns of AH, driven by the temperature differences across AC, are quite different between the standing and up-facing eyes. The temperature differences and the orientations of the eye have significantly influences on the flow patterns of AH, including the maximum velocity, the numbers of the vortex, the intensities and center positions of the vortex and so on, which may be beneficial to understand the formation of Krukenberg's Spindle.

Basing on the verification simulations, we systematically research the influences of the AH secretion rate, permeability of TM, viscosity of AH and anterior chamber angle on IOP in both healthy and pathologic eyes, the WSSes on the corneal endothelium and the iris surface, and the following conclusions can be drawn from the detailed analysis of the results.

(1) With the decrease of the inlet velocity or secretion rate of AH, IOP linearly reduces in both healthy and pathologic eyes. Furthermore, the IOP reduction exhibits a steeper gradient in pathologic eye than in healthy one. Therefore, through reducing the AH secretion rate with drugs, IOP may be expected to be controlled efficiently.

(2) With the decrease of the TM permeability, IOP will gradually increase with increasing velocity. In more detail, there would be a critical value of the TM permeability (about $5.33 \times 10^{-15}$ m$^2$), above which IOP increases slowly and below which IOP will increase rapidly. This finding is consistent with the clinical practice, in which the diagnosis of glaucoma does not simply rely on the presence of high IOP.

(3) With the increase of the AH viscosity, IOPs in both healthy and pathologic eyes will all rise at different velocities, and IOP rises faster in the pathologic eye than in the healthy one. However, the IOP increase is insensible to the viscosity of AH, compared to the permeability of TM and AH secretion rate, which may be one of the causes that the viscosity would not have been considered as a factor for controlling IOP.

(4) With the decrease of the anterior chamber angle, IOPs increase in both healthy and pathologic eyes with increasing velocities. For the IOP increase rate, there would be some critical value of the anterior chamber angle (about 15 ℃), above which IOP increases slowly and below which IOP increases dramatically. In addition, the IOP increase is faster in the pathologic eye than in the healthy one, which indicates that the narrowing of the anterior chamber angle produces a more significant effect in a pathologic situation.

(5) The WSSes in standing orientation is notable higher than those in up-facing orientation, and the temperature difference have significant influence on the cornea and iris WSSes, while the influence of inflow velocity, permeability of TM and AH viscosity are not obvious. The effect on WSSes with all these factors in pathologic eye is almost same as those in healthy eye. The WSSes in our cases are much lower than the detaching threshold value for endothelial cell, so it is unlikely that these WSSes could detach pigments from the anterior iris surface or endothelial cells from the corneal endothelium.

Due to the advantages of the lattice Boltzmann method in modeling complex fluid systems, the present model can be expected to become a good alternative tool for studying the AH dynamics and apply to the dynamics of AH more deeply and obtain some more interesting results.

## Acknowledgements

This work was supported by the National Natural Science Foundation of China (Grant Nos.11862003, 81860635, 12062005), the Project of Guangxi Natural Science Foundation (Grant Nos.2017GXNSFDA198038, 2018GXNSFAA281302), the Project for Promotion of Young and Middle-aged Teachers' Basic Scientific Research Ability in Guangxi Universities (Grant No. 2019KY0084), Guangxi Collaborative Innovation Center of Multi-source Information Integration and Intelligent Processing.

## Conflict of interest

The authors declare no competing interests.